\documentclass[twocolumn,amsmath,amssymb,aip,pop]{revtex4}

\usepackage{graphicx}
\usepackage{bm}

\usepackage{amsfonts}

\begin{document}
\title{Stable three-dimensional Langmuir vortex soliton }
\author{Volodymyr M. Lashkin}

 \affiliation{Institute for Nuclear
Research, Pr. Nauki 47, Kyiv 03680, Ukraine}

\email{vlashkin62@gmail.com}


\begin{abstract}
We present a numerical solution in the form of a three-dimensional
(3D) vortex soliton in unmagnetized plasma in the model of the
generalized Zakharov equations with saturating exponential
nonlinearity. To find the solution with a high accuracy we use
two-step numerical method combining the Petviashvili iteration
procedure and the Newton-Kantorovich method. The vortex soliton
with the topological charge $m=1$ turns out to be stable provided
the nonlinear frequency shift exceeds a certain critical value.
The stability predictions are verified by direct simulations of
the full dynamical equation.
\end{abstract}

\maketitle

\section{Introduction}

A vortex soliton (spinning soliton) is the localized nonlinear
structure with embedded vorticity and ringlike in the
two-dimnsional (2D) or toroidal in 3D case field intensity
distribution, with the dark hole at the center where the phase
dislocation takes place: a phase circulation around the azimuthal
axis is equal to $2m\pi$. An integer $m$ is referred to as
topological charge. The important integral of motion associated
with this type of solitary wave is the angular momentum. Spinning
solitons have attracted a great deal of attention primarily in
such fields of nonlinear physics as nonlinear optics and
Bose-Einstein condensates (BEC), where they are the subject of
considerable theoretical and experimental research (see a recent
review \cite{Malomed19} and also \cite{Kivshar_book} and
references therein). In models with cubic self-focusing
nonlinearity both fundamental (i. e. nonspinning, $m=0$) and
spinning solitons collapse (the amplitudes grow to infinity in a
finite time or propagation distance) in 2D and 3D dimensions.
Vortex solitons, unlike fundamental ones, in addition to the
collapse-driven instability, may undergo an even stronger
azimuthal instability, which tends to break the axially symmetric
ring or torus into fragments, each one being, roughly speaking, a
fundamental soliton. Stabilization of 2D and 3D vortex solitons
both against the collapse and azimuthal instability may be
achieved by means of competing nonlinearities, nonlocal
nonlinearities or by trapping potentials (harmonic-oscillator and
spatially-periodic ones) \cite{Malomed19}.

In plasmas, unlike optics and BEC, relatively few works have
addressed spinning solitons. Two-dimensional spinning solitons in
underdense plasma with relativistic saturating nonlinearity were
found in Ref. \cite{Berezhiani2002}. Those vortex solitons did not
collapse, but were unstable against azimuthal symmetry-breaking
perturbations that caused splitting of the rings into filaments
which form stable fundamental solitons. For electron-positron
plasmas with the temperature asymmetry of plasma species, 2D and
3D spinning solitons were studied in Ref. \cite{Berezhiani2010}.
The vanishing saturating nonlinearity in this case does not
sustain solitonic solutions with sufficiently large amplitudes in
contrast to the ordinary saturating nonlinearity \cite{Malomed19}.
Under this, 2D vortex solitons with amplitudes (or, equivalently,
nonlinear frequency shifts) below and above some critical value
turn out to be stable while the 3D ones split into filaments due
to the azimuthal instability. Stable 2D spinning solitons were
also found in partially ionized collision plasma with the so
called thermal nonlinearity \cite{Yakimenko05} and at the hybrid
plasma resonance \cite{Lashkin2007}. In both cases the stability
was due to the nonlocal character of the nonlinearity that is when
the nonlinear response depends on the wave packet intensity at
some extensive spatial domain and the nonlinear term has the
integral form. In Ref. \cite{Yakimenko05} the parameter of
nonlocality is related to the relative energy that the electron
delivers to the ion during single collision. It was shown that the
symmetry-breaking azimuthal instability of the 2D vortex soliton
with $|m|=1$ is fully eliminated in a highly nonlocal regime,
while the multicharge vortices with $|m|>1$ remain unstable with
respect to a decay into the fundamental solitons (driven-collapse
instability is absent for the all solitons). In Ref.
\cite{Lashkin2007} the nonlinear interaction between upper-hybrid
and dispersive magnetosonic waves was studied. Under this,
dispersion of the magnetosonic wave effectively introduces a
nonlocal nonlinear interaction and vortex solitons in this model
turn out to be stable if the amplitudes exceeds some critical
value.

To avoid misunderstanding it should be noted that in plasma
physics the term "vortex soliton" (or "solitary vortex") often
mean 2D \cite{Petviashvili92,Horton96} (for the only 3D case see
Ref. \cite{Lashkin2017}) dipole vortex structures with zero net
total angular momentum and these nonlinear structures (which are
studied in detail on some branches of the plasma oscillations
\cite{Petviashvili92}). Such dipole solitary vortices have nothing
to do with the spinning solitons.

The aim of the present work is to find a numerical solution in the
form of a stable 3D spinning soliton and demonstrate its stability
in unmagnetized plasma in the framework of the generalized
Zakharov equations with saturating exponential nonlinearity. This
type of nonlinearity is valid for sufficiently large field
amplitudes \cite{Kuznetsov86} and, in particular, is important for
the problem of inertial nuclear fusion \cite{Ruocco19}. In order
to find the numerical solution with a very high accuracy, we
present two-step numerical method combining the Petviashvili
iteration procedure \cite{Petviashvili76,Lakoba07} and the
Newton-Kantorovich method \cite{Kantorovich39,Kantorovich73}. We
show that such 3D vortex soliton is stable if the nonlinear
frequency shift exceeds a certain critical value and demonstrate,
by direct numerical simulations, that it can evolve without
distortion of the form over long time even under sufficiently
strong initial noise.

\section{Model equation}

In the simplest case of an unmagnetized plasma, dynamics of
nonlinear Langmuir waves is governed by the classical Zakharov
equations \cite{Zakharov72}. In the subsonic limit,  when
$\omega\ll kv_{s}$, where $\omega$ and $k$ are the characteristic
frequency and wave number of wave motions respectively,
$v_{s}=\sqrt{T_{e}/m}$ is the ion sound velocity, $T_{e}$ and $m$
are electron temperature and the electron mass respectively, the
Zakharov equations reduce to one equation
\begin{equation}
\label{Zakharov72} \Delta\left(i\frac{\partial\varphi}{\partial
t}+\frac{3}{2}\omega_{p}r_{D}^{2}
\Delta\varphi\right)-\frac{\omega_{p}}{2}\nabla\cdot
\left(\frac{\delta n}{n_{0}}\nabla\varphi\right)=0,
\end{equation}
for the slow varying complex amplitude $\varphi$ of the potential
of the electrostatic electric field $\mathbf{E}$
\begin{equation}
\label{E} \mathbf{E}=-\frac{1}{2}[\nabla\varphi\exp
(-i\omega_{p}t)+\mathrm{c.c.}]
\end{equation}
at the Langmuir frequency $\omega_{p}=\sqrt{4\pi e^{2}n_{0}/m}$,
where $e$ is the magnitude of the electron charge, $n_{0}$ is the
equilibrium plasma density and $r_{D}=T_{e}/4\pi e^{2} n_{0}$ is
the Debye length. Here, $\delta n$ is the plasma density
perturbation
\begin{equation}
\label{n3} \delta n=-\frac{|\nabla\varphi|^{2}}{16\pi n_{0}
T_{e}}.
\end{equation}
In the one-dimensional case Eqs. (\ref{Zakharov72}) and (\ref{n3})
for $E=-\partial\varphi/\partial x$ reduces to the well known
nonlinear Schr\"{o}dinger equation (NLSE). In 3D case, the cubic
nonlinearity in Eqs. (\ref{Zakharov72}) and (\ref{n3}) results in
the wave collapse, i.e. the situation where the wave amplitude
becomes singular within a finite time \cite{Zakharov72}. It should
be noted that Eqs. (\ref{Zakharov72}) and (\ref{n3}) in 2D and 3D
cases can not be reduced to the multidimensional scalar NLSE.
These equations, unlike 2D and 3D NLSE, contains the vector
differential operators. This difference, in particular, results in
the anisotropic form of the collapsing cavity, that is a region
with negative density $\delta n$ and trapped Langmuir waves
\cite{Kuznetsov86}. Furthermore, Eqs. (\ref{Zakharov72}) and
(\ref{n3}) for $E=-\partial\varphi /\partial r$ in radially
symmetric case contains an additional centrifugal term $((D
-1)/r^{2} )E$ , where $D$ is the space dimension, and, therefore,
the field $E\rightarrow 0$ as $r\rightarrow 0$.

The nonlinear dynamic motions corresponding to a collapse reach
dimensions of the order of $r_{D}$ where Landau damping occurs.
However, some experimental observations
\cite{Antipov81,Wong84,Cheung85} of 3D Langmuir collapse
demonstrate saturation of the field amplitude on a spatial scale
much larger compared to the Debye length $r_{D}$. Under this,
evolution of Langmuir wave packets shows slow dynamics with a
characteristic time scale $t\gg\omega_{pi}^{-1}$ (subsonic
regime), where $\omega_{pi}$ is the ion Langmuir frequency. From
the theoretical point of view, the wave collapse may be prevented
by including some extra effects such as higher order
nonlinearities, electron nonlinearities
\cite{Kuznetsov76,Scorich,Davydova2005}, saturating nonlinearity
\cite{Laedke84}, nonlocal nonlinearity \cite{Lashkin2007} etc..
Then, the arrest of collapse can result in the formation of
stationary structures which turn out to be (quasi)stable in some
regions of parameters. We consider the case of saturating
nonlinearity when the characteristic times of the nonlinear
processes to exceed significantly the time of an ion passing
through the cavity, and then both electrons in slow motions and
ions can be considered to have a Boltzmann distribution
\cite{Kaw1973,Kuznetsov86}
\begin{equation}
\label{n_Bolz} \frac{ \delta n}{n_{0}}=\exp
\left(-\frac{|\nabla\varphi|^{2}}{16\pi n_{0}T_{e}}\right)-1.
\end{equation}
Substituting Eq. (\ref{n_Bolz}) into Eq. (\ref{Zakharov72}) we get
\begin{gather}
\Delta\left(i\frac{\partial\varphi}{\partial
t}+\frac{3}{2}\omega_{p}r_{D}^{2} \Delta\varphi\right) \nonumber\\
-\frac{\omega_{p}}{2}\nabla\cdot \left\{\left[\exp
\left(-\frac{|\nabla\varphi|^{2}}{16\pi n_{0}T_{e}}\right)
 -1 \right]\nabla\varphi\right\}=0,
 \label{Zakharov_exp}
\end{gather}
Introducing the variables $\mathbf{r}^{'}$, $t^{'}$, $n^{'}$ and
$\varphi^{'}$ by
\begin{gather}
 \mathbf{r}=
\frac{3}{2}r_{D}\sqrt{\frac{M}{m}}\mathbf{r}^{'}, \,\, t=
\omega_{pe}^{-1}\sqrt{\frac{M}{m}} t^{'} , \nonumber \\
\frac{\delta n}{n_{0}}=\frac{4}{3}\frac{m}{M}n^{'}, \,\, \varphi=
\frac{T_{e}}{e}\sqrt{12}\varphi^{'} \label{dimensionless}
\end{gather}
we rewrite Eq. (\ref{Zakharov_exp}) in the dimensionless form
(accents have been omitted)
\begin{equation}
\label{eq1} \Delta\left(i\frac{\partial\varphi}{\partial t}+
\Delta\varphi\right)-\nabla\cdot \left\{[\exp
(-|\nabla\varphi|^{2})-1]\nabla\varphi\right\}=0.
\end{equation}
Equation (\ref{eq1}) can be written in Hamiltonian form
\begin{equation}
\label{hamil}  i\frac{\partial}{\partial t}\Delta\varphi
+\frac{\delta H}{\delta \varphi^{\ast}}=0 ,
\end{equation}
where the Hamiltonian is
\begin{equation}
\label{H} H=\int
\left[|\Delta\varphi|^{2}-|\nabla\varphi|^{2}-\exp
(-|\nabla\varphi|^{2})+1\right]d\mathbf{r}
\end{equation}
It follows immediately from (\ref{hamil}) that the Hamiltonian $H$
is conserved. Other integrals of motion are the plasmon number (a
consequence of the gauge invariance)
\begin{equation}
\label{N} N=\int |\nabla\varphi|^{2}d\mathbf{r},
\end{equation}
the momentum (a consequence of the translational invariance)
\begin{equation}
\label{P} \mathbf{P}=\int \mathbf{p}d\mathbf{r},
\end{equation}
where the momentum density is
\begin{equation}
p_{l}=\frac{i}{2}
(\nabla_{k}\varphi^{\ast}\nabla_{l}\nabla_{k}\varphi-\mathrm{c.
c.}),
\end{equation}
and the angular momentum (a consequence of the rotation
invariance)
\begin{equation}
\label{M} \mathbf{M}=\int
([\mathbf{r}\times\mathbf{p}]+i[\nabla\varphi\times\nabla\varphi^{\ast}])d\mathbf{r}.
\end{equation}
In the radially symmetric case stable three-dimensional soliton
solution of Eq. (\ref{Zakharov_exp}) was predicted in Ref.
\cite{Laedke84}.

Since the Langmuir field is a longitudinal one we have
$\mathbf{E}_{ \mathbf{k}}=(\mathbf{k}/k)E_{\mathbf{k}}$, where
$\mathbf{E}_{\mathbf{k}}$ is the space Fourier transform of the
electric field $\mathbf{E}$ ($\mathbf{k}$ is the wave vector,
$k\equiv|\mathbf{k}|$) and next consider the magnitude $E$ instead
of the potential $\varphi$. In the Fourier space equation
(\ref{eq1}) can be written as
\begin{equation}
\label{eq11} i\frac{\partial\varphi_{\mathbf{k}}}{\partial
t}-k^{2}\varphi_{\mathbf{k}} -\frac{\mathbf{k}}{k^{2}}\cdot\int
n_{\mathbf{k}_{1}}\mathbf{k}_{2}\varphi_{\mathbf{k}_{2}}\delta
(\mathbf{k}-\mathbf{k}_{1}-\mathbf{k}_{2})d\mathbf{k}_{1}d\mathbf{k}_{2}=0,
\end{equation}
where $n_{\mathbf{k}}$ is the Fourier transform of the density
perturbation determined by Eq. (\ref{n_Bolz}). On the other hand,
the Fourier transform of $|\nabla\varphi|^{2}$ in Eq.
(\ref{n_Bolz}) is
\begin{equation}
\label{n222}
|\nabla\varphi|_{\mathbf{k}}^{2}=-\int\mathbf{k}_{1}\cdot\mathbf{k}_{2}
\varphi_{\mathbf{k}_{1}}\varphi_{\mathbf{k}_{2}}^{\ast}\delta
(\mathbf{k}-\mathbf{k}_{1}-\mathbf{k}_{2})d\mathbf{k}_{1}d\mathbf{k}_{2}.
\end{equation}
By introducing the operator $\hat{\mathbf{L}}$ acting in the
physical space as ($f(\mathbf{r})$ is the arbitrary function and
$\hat{f}(\mathbf{k})$ is its Fourier transform)
\begin{equation}
\hat{\mathbf{L}}f(\mathbf{r})=\int
\hat{f}(\mathbf{k})\frac{\mathbf{k}}{k}\mathrm{e}^{-i\mathbf{k}\cdot\mathbf{r}}d\mathbf{k}
\end{equation}
and taking into account that
$E_{\mathbf{k}}=-ik\varphi_{\mathbf{k}}$, Eqs. (\ref{eq11}) and
(\ref{n222})  become
\begin{equation}
\label{eq222} i\frac{\partial E_{\mathbf{k}}}{\partial t}-k^{2}
E_{\mathbf{k}}+\hat{\mathbf{L}}_{\mathbf{k}}\cdot \int
n_{\mathbf{k}_{1}}\hat{\mathbf{L}}_{\mathbf{k}_{2}}E_{\mathbf{k}_{2}}\delta
(\mathbf{k}-\mathbf{k}_{1}-\mathbf{k}_{2})d\mathbf{k}_{1}d\mathbf{k}_{2}=0,
\end{equation}
and
\begin{equation}
\label{eq2222} |\nabla\varphi|_{\mathbf{k}}^{2}=\int
\hat{\mathbf{L}}_{\mathbf{k}_{1}}E_{\mathbf{k}_{1}}\cdot
\hat{\mathbf{L}}_{\mathbf{k}_{2}}E_{\mathbf{k}_{2}}\delta
(\mathbf{k}-\mathbf{k}_{1}-\mathbf{k}_{2})d\mathbf{k}_{1}d\mathbf{k}_{2},
\end{equation}
respectively. Then, in the physical space one can write Eq.
(\ref{eq222}) in the form
\begin{equation}
\label{eq2} i\frac{\partial E}{\partial t}+\Delta
E+\hat{\mathbf{L}}\cdot (n\hat{\mathbf{L}}E)=0,
\end{equation}
where
\begin{equation}
\label{eq22} n=\exp (-|\hat{\mathbf{L}}E|^{2})-1.
\end{equation}

\section{Vortex soliton solution}

We look for a stationary solution of  (\ref{eq2}) in the form
\begin{equation}
\label{lambd} E(\mathbf{r},t)=A(\mathbf{r})\exp (i\lambda t),
\end{equation}
where $\lambda$ is the nonlinear frequency shift. Substituting
(\ref{lambd}) into (\ref{eq2}) and (\ref{eq22}), one can obtain
\begin{equation}
\label{eq4} -\lambda A+\Delta A+\hat{\mathbf{L}}\cdot
(n\hat{\mathbf{L}}A)=0,
\end{equation}
where
\begin{equation}
\label{eq44} n=\exp (-|\hat{\mathbf{L}}A|^{2})-1.
\end{equation}
We are interested in the stationary solution in the form of a
solitary vortex with axial symmetry
\begin{equation}
\label{axial} A(\mathbf{r})=\mathcal{A}(r,z)\exp (im\theta)\equiv
\mathcal{A}(r,z)\frac{(x\pm iy)^{|m|}}{r^{|m|}},
\end{equation}
where $r=\sqrt{x^{2}+y^{2}}$ and $\theta$  are  the radial
coordinate and the azimuthal angle, respectively, in the
cylyndrical coordinates $(r,\theta,z)$. The real function
$\mathcal{A}(r)$ should satisfy the boundary conditions at the
centre, $\mathcal{A}\rightarrow 0$ as $r\rightarrow 0$ and at
infinity, $\mathcal{A}\rightarrow 0$ as $r\rightarrow\infty$. An
integer $m$ referred as the topological charge. The signs $\pm$
correspond to $\pm |m|$. Such solutions describe either the
fundamental soliton, when $m=0$, or the vortex soliton with the
topological charge $m\neq 0$. In contrast to the case $m=0$, one
can see that due to the vector type of the nonlinearity it is
difficult to write the equation for the function
$\mathcal{A}(r,z)$ if $m\neq0$. Thus, we must solve equations
(\ref{eq4}), (\ref{eq44})  in cartesian coordinates $(x,y,z)$ just
for the function $A(\mathbf{r})$ without specifying the vortex
topology. The principal difficulty for finding numerical solutions
of the vortex type is the convergence of usual relaxation
iteration procedures to the ground state, i. e. to the fundamental
soliton rather than the vortex soliton (or any other excited
states). To this end, we use two stage method which combines the
Petviashvili iteration procedure (the first step)
\cite{Petviashvili76,Lakoba07} and the Newton-Kantorovich method
(the second step) \cite{Kantorovich39,Kantorovich73}. The latter,
as it is known, belongs to a family of universally convergent
iterative methods and it can converge to any nonfundamental
solution provided that the initial condition is sufficiently close
to that solution. In the $\mathbf{k}$-space equations (\ref{eq4})
and (\ref{eq44}) can be written as
\begin{equation}
\label{eq5} -\lambda A_{\mathbf{k}}-k^{2}A_{\mathbf{k}}=\int
\frac{\mathbf{k}\cdot\mathbf{k}_{2}}{kk_{2}}n_{\mathbf{k}_{1}}A_{\mathbf{k}_{2}}\delta
(\mathbf{k}-\mathbf{k}_{1}-\mathbf{k}_{2})d\mathbf{k}_{1}d\mathbf{k}_{2},
\end{equation}
where $n_{\mathbf{k}}$ is the Fourier transform of the plasma
density perturbation $n$ determined by (\ref{eq44}) and
\begin{equation}
\label{eq444} |\hat{\mathbf{L}}A|^{2}_{\mathbf{k}}=\int
\frac{\mathbf{k}_{1}\cdot\mathbf{k}_{2}}{k_{1}k_{2}}A_{\mathbf{k}_{1}}A_{\mathbf{k}_{2}}^{\ast}\delta
(\mathbf{k}-\mathbf{k}_{1}-\mathbf{k}_{2})d\mathbf{k}_{1}d\mathbf{k}_{2}.
\end{equation}
\begin{figure}
\includegraphics[width=3.4in]{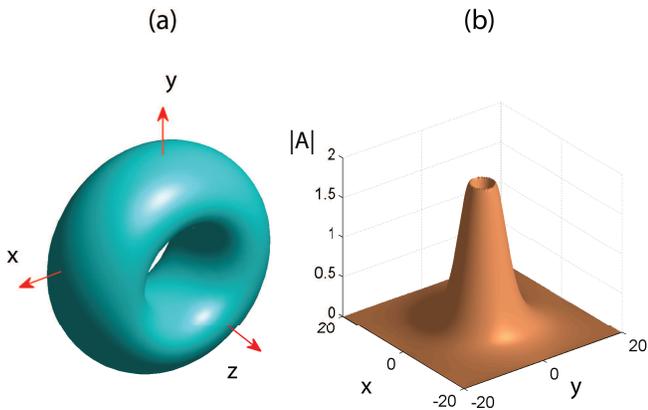}
\caption{\label{fig1}  The vortex soliton with $\lambda=0.2$. Left
column: isosurface $|A(x,y,z)|=0.15$; right column: the field
$|A|$ in the $x-y$ plane (i.e. in vertical cross-section, $z =
0$).}
\end{figure}
Note that the nonlinearity in (\ref{eq5}), (\ref{eq444}) has an
explicitly anisotropic character. To calculate nonlinear terms in
physical and Fourier spaces we use the identity ($f$ and $g$ are
arbitrary functions)
\begin{equation}
(fg)_{\mathbf{k}}=\int\hat{f}_{\mathbf{k_{1}}}\hat{g}_{\mathbf{k_{2}}}\delta
(\mathbf{k}-\mathbf{k}_{1}-\mathbf{k}_{2})\,d\mathbf{k}_{1}d\mathbf{k}_{2}.
\end{equation}
Equation (\ref{eq5}) can be written in the form
\begin{equation}
\label{eq00} G_{\mathbf{k}}A_{\mathbf{k}}=B_{\mathbf{k}},
\end{equation}
where $G_{\mathbf{k}}=-(\lambda+k^{2})$ and $B_{\mathbf{k}}$
accounts for the nonlinear term.  The Petviashvili iteration
method for solving Eq. (\ref{eq00}) is presented in the Appendix.
An initial guess is chosen in the form of the vortex soliton
(\ref{axial}) with $\mathcal{A}(r,z)=\sqrt{\lambda}r^{|m|}\exp
[-\sqrt{\lambda}(r^{2}+z^{2})]$. Next we restrict the case $m=1$.
The convergence was controlled by stopping the iteration when the
value $|s-1|$ began to increase, where $s$ is determined by Eq.
(\ref{s}). This indicates that the iteration procedure jumps off
the solution corresponding the vortex soliton and begins to
converge to the fundamental soliton. Under this, the found
approximate vortex solution has a quite acceptable accuracy:
typically, depending on $\lambda$, one can reach the value
$|s-1|\sim 10^{-2}-10^{-3}$. Then, as the second step, this
obtained solution is used as an initial condition in the
Newton-Kantorovich iterative method as described in the Appendix.
\begin{figure}
\includegraphics[width=3.in]{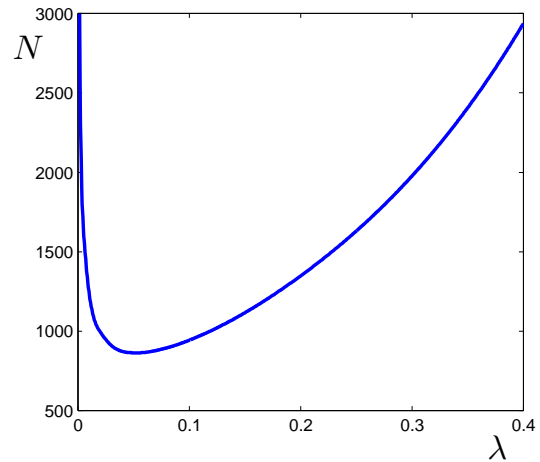}
\caption{\label{fig2}  The plasmon number $N$ of the 3D vortex
soliton as a function of the nonlinear frequency shift $\lambda$.
}
\end{figure}
The method reduces the corresponding nonlinear problem into a
sequence of linear equations (\ref{Kantor}) which are solved by
the conjugate gradient method \cite{Dongarra1994}. The
Newton-Kantorovich method has a quadratic rate of convergence and
typically after several iterations we are able to find the vortex
soliton solutions with a very high accuracy (up to a machine one).
An example of the vortex soliton solution with $\lambda=0.2$ is
presented in Fig.~\ref{fig1}.

\begin{figure}
\includegraphics[width=3.4in]{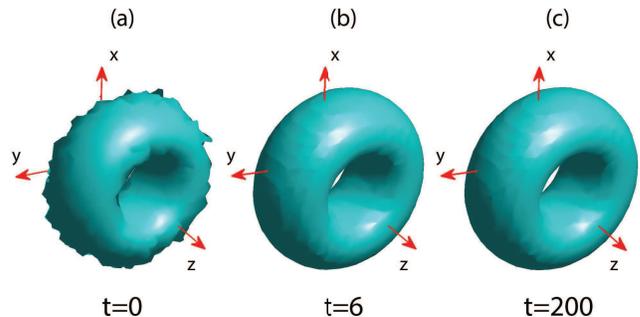}
\caption{\label{fig3}  Self-cleaning of a randomly perturbed
stable vortex soliton with $m=1$ after the application of a random
perturbation with the parameter $\nu=0.08$. Panels (a), (b) and
(c) display the shape of the perturbed vortex at the initial
moment, $t = 0$, at $t=6$ and at $t=200$. The unperturbed vortex
has $\lambda=0.2$ and $N =1347$. }
\end{figure}
The plasmon number (\ref{N}) of the vortex soliton given by the
expression
\begin{equation}
\label{N1} N=\int |\hat{\mathbf{L}}A|^{2}d\mathbf{r}
\end{equation}
is plotted, as a function of the nonlinear frequency shift
$\lambda$, in Fig.~\ref{fig2}. The minimum of the function,
$N_{cr}=863$, is located at $\lambda_{cr}=0.053$.

The physical relevance of any stationary solution depends on
whether it is stable. A well known approach to predicting the
stability for soliton families in NLSE of any dimension is based
on the Vakhitov-Kolokolov (VK) criterion
\cite{Petviashvili92,Vakhitov73}. The VK criterion states that a
stationary solution $\sim \exp (i\lambda t)$ with the energy $N$
may be stable if $\partial N/\partial \lambda>0$, and is
definitely unstable otherwise. For the fundamental solitons (i. e.
ground states), this condition is both a necessary and sufficient
one. This criterion, however, does not apply to the equation
(\ref{eq1}) and moreover, provides a necessary, but generally, not
sufficient condition for the stability of vortex solitons. In the
region where the VK criterion holds, but the vortices in NLSE are
unstable, they are vulnerable to the splitting instability induced
by perturbations breaking the azimuthal symmetry \cite{Malomed19}.

To investigate the vortex soliton stability within the framework
of Eq. (\ref{eq1}) we solved numerically the dynamical equations
(\ref{eq2}) and (\ref{eq22}) initialized with our computed vortex
solutions with added initial perturbation. The initial condition
was taken in the form $E(\mathbf{r},0)=A_{0}(\mathbf{r})[1+\nu
f(\mathbf{r})]$, where $A_{0}(\mathbf{r})$ is the numerically
calculated solution and $f(\mathbf{r})$ is some function
corresponding the perturbation at the initial time $t=0$. We
considered two forms of the function $f(\mathbf{r})$. In the first
case, $f(\mathbf{r})$ is the white Gaussian noise with the zero
mean and variance $\sigma^{2}=1$. In the second case,
$f(\mathbf{r})=\sin x+i\cos y$. The parameter of perturbation is
$\nu=0.005-0.1$. In both cases perturbations break the azimuthal
symmetry. A numerical simulation shows that the vortex soliton
turns out to be stable if $\lambda>\lambda_{cr}$ that is in the
region formally predicted by VK criterion. We could not see any
evidence of the splitting instability at least up to times $t=400$
that is much larger than the characteristic nonlinear time $\sim
1/\lambda$. An example of stable evolution of the vortex soliton
with $\lambda=0.2$ is shown in Fig.~\ref{fig3}. The initial state
of the vortex soliton is perturbed by a rather strong noise with
$\nu=0.08$. The vortex soliton cleans up itself from the noise at
the characteristic nonlinear time $\sim 1/\lambda$ and then
survives over long time.

Though we did not perform the linear stability analysis with the
corresponding eigenvalue problem, it seems that the complex
eigenvalues (if any) accounting for the splitting instability are
sufficiently small. In this connection it is interesting to
compare some results concerning the stability of the
vortex-soliton solutions obtained for the generalized
two-dimensional NLSE of the form
\begin{equation}
\label{NLS2D_general} i\frac{\partial\psi}{\partial
t}+\Delta_{\bot}\psi+f(|\psi|^{2})\psi=0,
\end{equation}
where $f(|\psi|^{2})$ is an arbitrary function of the intensity
and $\Delta_{\bot}=\partial^{2}/\partial
x^{2}+\partial^{2}/\partial y^{2}$. Following the approach
originally proposed in Refs. \cite{Soto-Crespo92,Soto-Crespo94},
an approximate analytical estimate for the growth rate of
azimuthal instability against the azimuthal perturbations
$\sim\exp (i\Omega t+iM\theta)$ of the vortex soliton in the model
(\ref{NLS2D_general}) can be written as \cite{Berge2006}
\begin{equation}
\label{gamma_Crespo}
\mathrm{Im}\,\Omega=\frac{M}{\bar{r}_{m,\lambda}}\mathrm{Re}\left[
2f^{'}(|\psi_{0}|^{2})|\psi_{0}|^{2}-\frac{M}{\bar{r}_{m,\lambda}^{2}}\right]^{1/2}
\end{equation}
Here the prime stands for the derivative, $\bar{r}_{m,\lambda}$ is
the mean value of the vortex radius defined in Ref.
\cite{Soto-Crespo92} and the amplitude
$|\psi_{0}|\equiv|\psi_{m,\lambda}|$ is evaluated at
$r=\bar{r}_{m,\lambda}$ ($m$ is the topological charge and
$\lambda$ is the nonlinear frequency shift). For the competing
nonlinearity $f(I)=I-I^{2}$ in Eq. (\ref{NLS2D_general}), where
$I=|\psi|^{2}$, the function $f^{'}(I)$ may be negative so that
$\mathrm{Im}\,\Omega=0$ and the vortex turns out to be stable for
sufficiently large amplitudes \cite{Pego2002}. For the saturating
nonlinearity with $f(I)=I/(1+I)$ the function $f^{'}(I)$ is always
positive and spinning solitons are unstable in agreement with the
rigorous analysis \cite{Skryabin1998}. In the case of the
exponential saturating nonlinearity $f(I)=1-\exp (-I)$, which has
not been previously studied, the first term in the square brackets
in Eq. (\ref{gamma_Crespo}), though positive, is exponentially
small for large amplitudes and the instability is either
practically suppressed or absent. In our case, by analogy with the
model (\ref{NLS2D_general}) one could also expect strong
decreasing of the growth rate or full stability of the vortex
soliton.

It is important to emphasize that at the critical value
$\lambda_{cr}$ the vortex soliton characteristic size
$\sim\lambda_{cr}^{-1/2}$ is much larger than the electron Debye
length so that the Landau damping is negligible. The Landau
damping is still small enough for the solitons with
$\lambda\lesssim 0.2$, however for larger $\lambda$ values one has
to take into account the damping.

While the fundamental soliton corresponds to the ground state of
the nonlinear eigenvalue problem, the vortex soliton may be
regarded as an excited state. Under certain conditions, one can
expect emergence of the excited states (sometimes with a
sufficiently long lifetime) just as in linear problems. For
example, dynamical generation of the vortex soliton clusters from
the initial condition in the form of the fundamental soliton with
superimposed discontinuous phases was demonstrated in the
two-dimensional NLSE with the parabolic trapping potential
\cite{Lashkin2012}.

\section{Conclusion}

In conclusion, we have shown the possibility of existence in an
unmagnetized plasma of the stable 3D solitons with the nonzero
angular momentum (vortex solitons). Such coherent structures, as
well as fundamental solitons, may be considered as "elementary
bricks" of strong Langmuir turbulence. The real picture of strong
Langmuir turbulence -- collapsing cavitons or quasistationary
structures interacting with quasilinear waves -- is apparently
largely dependent on excitation conditions, amplitudes and spatial
scales of initial perturbations.

\appendix*
\section{Petviashvili and Newton-Kantorovich iteration schemes}

The Petviashvili iteration procedure
\cite{Petviashvili76,Lakoba07} for Eq. (\ref{eq00}) at the $n$-th
iteration is
\begin{equation}
A_{\mathbf{k}}^{(n+1)}=sG_{\mathbf{k}}^{-1}B_{\mathbf{k}}^{(n)},
\end{equation}
where $s$ is the so called stabilizing factor defined by
\begin{equation}
\label{s} s=\left(\frac{\int
|A_{\mathbf{k}}^{(n)}|^{2}d\mathbf{k}}{\int
A_{\mathbf{k}}^{\ast,(n)}G_{\mathbf{k}}^{-1}B_{\mathbf{k}}^{(n)}
d\mathbf{k}}\right)^{\alpha}.
\end{equation}
and $\alpha >1$. For the power nonlinearity, the fastest
convergence is achieved for $\alpha=p/(p-1)$, where $p$ is the
power of nonlinearity \cite{Lakoba07}. For the exponential
nonlinearity in (\ref{eq44}) we use, depending on $\lambda$,
empirical value $\alpha=1.2-1.3$.

The equation (\ref{eq4}) is rewritten in the form
$F(\mathbf{A})=0$ with $\mathbf{A}=(A,A^{\ast})$). Then
Newton-Kantorovich iteration scheme
\cite{Kantorovich39,Kantorovich73} is
\begin{equation}
\label{Kantor}
F^{'}(\mathbf{A}^{(n)})\mathbf{A}^{(n+1)}=F^{'}(\mathbf{A}^{(n)})\mathbf{A}^{(n)}
-F(\mathbf{A}^{(n)}),
\end{equation}
where the prime stands for the Frechet derivative of the operator
$F(\mathbf{A})$ at the point $\mathbf{A}_{0}$ defined as
\begin{equation}
F^{'}(\mathbf{A}_{0})\mathbf{A}=\lim_{h\rightarrow
0}[F(\mathbf{A}_{0}+h\mathbf{A})-F(\mathbf{A}_{0}))].
\end{equation}
Calculating the Frechet derivative reduces to linearizing the
corresponding nonlinear operator in $h$ and
$F^{'}(\mathbf{A}^{(n)})$ is determined by
\begin{equation}
F^{'}(\mathbf{A}^{(n)})h=-\lambda h+\Delta h+\hat{\mathbf{L}}\cdot
(\alpha\hat{\mathbf{L}}h-\beta\hat{\mathbf{L}}h^{\ast}),
\end{equation}
where
\begin{equation}
\label{alpha} \alpha=[\exp
(-|\hat{\mathbf{L}}A^{(n)}|^{2})-1]-\exp
(-|\hat{\mathbf{L}}A^{(n)}|^{2})|\hat{\mathbf{L}}A^{(n)}|^{2}
\end{equation}
and
\begin{equation}
\label{beta} \beta=\exp
(-|\hat{\mathbf{L}}A^{(n)}|^{2})(\hat{\mathbf{L}}A^{(n)})^{2}.
\end{equation}

\end{document}